\documentclass[a4paper,fleqn]{cas-sc}

\usepackage[numbers,sort&compress]{natbib}
\usepackage{float}

\begin{document}
\let\WriteBookmarks\relax

\shorttitle{Strategies for Molecular Dynamics using Hybrid Systems}

\shortauthors{P. H. L. Ramalho et al.}

\title[mode = title]{Strategies for Molecular Dynamics using Hybrid Systems: LAMMPS Use Case}


\author[1]{Paulo Henrique Leme Ramalho}[orcid=0009-0003-9870-841X]
\cormark[1]
\ead{paulo.henrique.ramalho@mail.usf.edu.br}
\credit{Conceptualization, Methodology, Software, Investigation, Writing -- original draft}
\affiliation[1]{organization={Universidade S\~ao Francisco (USF)},
            city={Bragan\c{c}a Paulista},
            state={S\~ao Paulo},
            country={Brazil}}

\author[2]{Dennis Alves Pedersen}[orcid=0009-0007-1864-0101]
\ead{dennis.pedersen@mail.usf.edu.br}
\credit{Software, Validation, Data curation, Writing -- review \& editing}
\affiliation[2]{organization={Universidade S\~ao Francisco (USF)},
            city={Itatiba},
            state={S\~ao Paulo},
            country={Brazil}}

\author[3]{F\'abio Andrijauskas}[orcid=0000-0002-1254-8570]
\ead{fabio.andrijauskas@usf.edu.br}
\credit{Conceptualization, Supervision, Resources, Writing -- review \& editing}
\affiliation[3]{organization={Universidade S\~ao Francisco (USF)},
            state={S\~ao Paulo},
            country={Brazil}}

\cortext[1]{Corresponding author}


\begin{abstract}
The complexity of biomolecular simulations has substantially increased the demand for High-Performance Computing (HPC) infrastructures, particularly in applications involving molecular dynamics and coarse-grained modeling of biological systems. The efficiency of parallelization strategies is a critical factor for the computational feasibility of simulations executed on modern multi-node architectures. This work presents a systematic performance and scalability analysis of the LAMMPS simulator for coarse-grained biomolecular simulations, using the antimicrobial peptide Tritrpticin (PDB ID: 1D6X) as the experimental workload. Equivalent scalability characterization for LAMMPS coarse-grained workloads on modern NUMA platforms has not previously appeared in the literature. Pure MPI and hybrid MPI+OpenMP executions were evaluated in HPC environments comprising up to 8 compute nodes and 1024 simultaneous cores. The experiments investigated metrics of execution time, speedup, parallel efficiency, statistical variability, and LAMMPS internal time decomposition. The results showed that pure MPI executions deliver excellent performance in single-node environments but suffer progressive scalability degradation in multi-node executions due to the increase in distributed communication overhead and inter-process synchronization. In contrast, hybrid MPI+OpenMP configurations proved more efficient at large scale, reducing communication costs and better exploiting the NUMA memory hierarchy of the underlying architecture. The analysis of the computational breakdown revealed that routines associated with communication and electrostatic interactions accounted for the largest fraction of total execution time at the largest pure-MPI scales. The results obtained reinforce the idea that the performance of biomolecular applications in HPC environments depends directly on the balance among parallelization granularity, spatial decomposition, and distributed communication costs. Thus, hybrid MPI+OpenMP strategies represent a more sustainable alternative for coarse-grained biomolecular simulations executed on modern many-core architectures.
\end{abstract}


\begin{highlights}
\item Pure MPI collapses at multi-node scale: Kspace+Comm reaches 89\% at 1024 cores
\item Hybrid $128\times8$ on 8 nodes recovers near single-node performance (4.51 s vs 4.40 s)
\item Working set fits in EPYC L3 cache; bottleneck is PPPM sync, not memory bandwidth
\item NUMA-aware MPI placement preserves CCD-local L3 residency and intra-node locality
\item Verification scripts cross-check every numeric claim against raw LAMMPS logs
\end{highlights}



\maketitle

\section{Introduction}\label{sec:intro}

Investigating the behavior of biomolecules at the atomic level has become feasible thanks to advances in molecular dynamics methods. This approach is today indispensable for understanding processes such as protein--ligand interaction and the design of new drugs. However, the practical application of these techniques faces performance bottlenecks, given that simulating systems with millions of atoms over extensive time scales demands extremely high computational power \citep{liuApplyingHighperformanceComputing2016,hollingsworthMolecularDynamicsSimulation2018}.

Given the computational complexity of these systems, the support of High-Performance Computing (HPC) is what makes research feasible. The transition to parallel environments allows previously impractical calculations to be completed within acceptable time windows. Among the available simulators, LAMMPS (Large-scale Atomic/Molecular Massively Parallel Simulator) stands out for its architecture, which is compatible with the main industry standards, such as distributed memory (MPI) and shared memory (OpenMP), offering the versatility needed to handle this complexity. While hybrid MPI+OpenMP scalability has been systematically characterized for simulators such as GROMACS and NAMD on large-scale HPC architectures \citep{acunScalableMolecularDynamics2018,kutznerScalingGROMACSMolecular2025}, equivalent analysis for LAMMPS coarse-grained biomolecular workloads --- particularly those using the SPICA force field on modern NUMA architectures --- remains absent from the literature. This gap motivates the present study. Critically, small- to medium-sized coarse-grained systems of this scale represent a widespread and practically relevant HPC workload class: many membrane--peptide interaction studies, lipid bilayer patch simulations, and initial equilibration stages of larger systems operate in precisely this particle-count range, yet their scalability characteristics on modern multi-node architectures have not been systematically documented for LAMMPS.

The choice of parallelism model, however, remains a critical point. Since performance and hardware utilization vary widely across techniques, a comparative analysis is essential to optimize cluster utilization. In the healthcare field, this saving of resources is strategic, since processing cost often limits the reach of investigations \citep{kochHPCMedicalField2023}.

Considering this, the present work analyzes LAMMPS performance in an HPC environment, comparing pure MPI and MPI+OpenMP hybrid executions on multi-node architectures. To this end, a protocol was established based on the calibration of a representative workload, followed by tests that evaluate metrics such as execution time, speedup, parallel efficiency, and memory consumption.

The central hypothesis is that, for small- to medium-sized coarse-grained biomolecular workloads, pure MPI executions will exhibit progressive scalability degradation at multi-node scale due to the growth of inter-process communication overhead, while hybrid MPI+OpenMP configurations will better exploit the NUMA hierarchy of modern multi-chiplet processors and sustain performance across node boundaries. The optimal granularity is expected to depend on the balance between spatial decomposition, inter-node communication cost, and thread-level parallelism. Validating this premise helps define more intelligent practices in the use of HPC, providing technical input that makes biomolecular simulations on shared infrastructure more agile and less costly.

\section{Theoretical background}\label{sec:background}

High-Performance Computing (HPC) has established itself as the technological foundation for contemporary scientific progress, enabling simulations across scientific domains previously inaccessible \citep{liuApplyingHighperformanceComputing2016}. HPC acts as an interdisciplinary catalyst, integrating computational chemistry and molecular biology to sustain scientific innovation \citep{liuApplyingHighperformanceComputing2016}. In the biomedical domain, the growth of genomic, imaging, and clinical data has reached volumetric levels comparable to astronomy \citep{stephensBigDataAstronomical2015}, imposing critical bottlenecks that render conventional computing obsolete and reinforce the urgency of HPC infrastructures \citep{lightbodyReviewApplicationsHighthroughput2019}. The evolution of distributed programming models such as MapReduce \citep{deanMapReduceSimplifiedData2008} and the democratization of scalable architectures \citep{gothEricBrewerChange2010} established the foundations of modern HPC, while genomic centers routinely generating 10--12 TB per week demand local high-bandwidth infrastructure \citep{kovatchOptimizingHighPerformanceComputing2020}. The strategic application of HPC in healthcare has enabled qualitative leaps in genomic analysis, pathology modeling, and clinical decision support \citep{liHighperformanceComputingHealthcare2024}, with the emerging HPC+ paradigm combining HPC, High-Performance Data Analytics, and AI to maximize the fidelity of complex simulations \citep{kochHPCMedicalField2023}. This convergence also drives machine-learning-guided molecular modeling and predictive computational medicine \citep{mashayakRelativeEntropyOptimizationDriven2015}.

In pharmacology, HPC enables molecular simulations of very high computational cost \citep{liuApplyingHighperformanceComputing2016}, with virtual screening and molecular docking methodologies reducing the timeline for drug development \citep{liuApplyingHighperformanceComputing2016}. Parallel pipelines have delivered substantial gains in research on neurodegenerative diseases such as Alzheimer's \citep{alliataParallelizingDrugDiscovery2025}, while coarse-grained modeling techniques reduce the computational burden of extensive biomolecular systems \citep{marrinkPerspectiveMartiniModel2013}.

\subsection{Molecular dynamics and coarse-grained models}\label{sec:md-cg}

Molecular Dynamics (MD) is among the most robust applications of HPC in structural biology, enabling the temporal observation of biomolecules to study protein--ligand interactions and conformational fluctuations \citep{hollingsworthMolecularDynamicsSimulation2018}. MD simulations operate by numerically integrating Newton's equations of motion to map atomic trajectories, from which structural and thermodynamic properties are derived \citep{braunBestPracticesFoundations2018}. This analytical capability is particularly valuable for phenomena inaccessible to direct experimentation, such as protein loop dynamics that determine inhibitor effectiveness \citep{kochHPCMedicalField2023}. The relevance of MD extends across the study of proteins and lipids whose functions depend on conformational rearrangements \citep{georgievaProteinConformationalDynamics2020}, gene regulatory networks in complex diseases \citep{ramirezModelingContextSpecificEMT2020}, rational drug design targeting proteases such as BACE1 in Alzheimer's disease \citep{herberClickChemistrymediatedBiotinylation2018}, and the optimization of therapies such as FLASH radiotherapy \citep{adrianFLASHEffectDepends2019}.

In complex biological systems --- particularly biomembranes and multicomponent lipid assemblies --- simulations require explicit representation of solvent, ions, and membrane components, significantly increasing particle count and computational cost \citep{seoSPICAForceField2018}. Coarse-grained (CG) approaches address this by grouping multiple atoms into effective particles (beads), reducing the system's degrees of freedom and enabling exploration of temporal and spatial scales inaccessible to fully atomistic models \citep{marrinkPerspectiveMartiniModel2013,seoSPICAForceField2018}. Force fields such as MARTINI and SPICA have become references for lipid membranes, molecular self-organization, and protein--membrane interactions \citep{seoSPICAForceField2018,monticelliForceFieldsClassical2012}. However, the massive computational cost of representing high-dimensional biological systems --- often millions of atoms with complex non-covalent interaction networks --- demands exponentially greater hardware resources \citep{hollingsworthMolecularDynamicsSimulation2018,braunBestPracticesFoundations2018}. Optimized simulation engines such as GROMACS, NAMD, and LAMMPS address this through hybrid execution models combining MPI and OpenMP, as well as GPU acceleration capable of speeds up to 150 times greater than CPU-only implementations \citep{kochHPCMedicalField2023,acunScalableMolecularDynamics2018}.

\subsection{Force fields and computational cost}\label{sec:forcefields}

The fidelity of an MD simulation lies in the mathematical description of interparticle forces consolidated in force fields, which act as a ``computational microscope'' providing resolution unattainable by static experimental methods \citep{drorBiomolecularSimulationComputational2012,lindorff-larsenSystematicValidationProtein2012}. The total potential energy combines bonded interactions (bond stretching, valence angles, torsion) and non-bonded interactions --- the dominant computational cost --- including Coulomb electrostatics and Lennard-Jones forces that describe the balance between short-range repulsion and long-range dispersion \citep{gelpiMolecularDynamicsSimulations2015,vanommeslaegheCHARMMGeneralForce2009}. Coarse-grained force fields such as MARTINI and SPICA preserve essential physicochemical fidelity while significantly reducing computational expense, though their parameterization requires careful balancing of efficiency and accuracy \citep{marrinkPerspectiveMartiniModel2013,seoSPICAForceField2018,monticelliForceFieldsClassical2012}. The validation of these potentials against experimental data enables correlation of atomic-scale anomalies with pathological outcomes, such as OPA1 mutations leading to mitochondrial coupling defects \citep{nochezAcuteLateonsetOptic2009}, and extends MD applicability to characterization of biocompatible biopolymers for medical devices \citep{righettiConstrainedAmorphousInterphase2019}. The selection of force fields combined with efficient parallel implementation ensures faithful reproduction of biomolecular structural dynamics \citep{georgievaProteinConformationalDynamics2020,monticelliForceFieldsClassical2012}.

Without optimization, non-bonded force evaluation scales as $O(N^2)$, but spatial decomposition and neighbor lists --- as implemented in LAMMPS --- reduce short-range complexity to $O(N)$, while long-range electrostatics are treated via Particle Mesh Ewald methods scaling as $O(N\log N)$ \citep{plimptonFastParallelAlgorithms1995,essmannSmoothParticleMesh1995}. The deep disparity between femtosecond integration steps and millisecond-scale biological processes imposes additional computational burden \citep{stephensBigDataAstronomical2015,gelpiMolecularDynamicsSimulations2015}, with even optimized simulations requiring weeks of continuous HPC execution for protein folding studies \citep{liuApplyingHighperformanceComputing2016}. Coarse-grained approaches have become fundamental to bridge this gap \citep{marrinkPerspectiveMartiniModel2013,seoSPICAForceField2018}.

\subsection{LAMMPS and parallelization challenges}\label{sec:lammps}

LAMMPS stands out as one of the most versatile MD codes, with a modular object-oriented architecture supporting atomic, mesoscopic, and continuum scales \citep{thompsonLAMMPSFlexibleSimulation2022,rohskopfFitSNAPAtomisticMachine2023}. Its extensibility enables incorporation of specialized CG force fields such as SPICA, supporting simulations involving millions of particles in distributed environments \citep{seoSPICAForceField2018}. LAMMPS implements spatial decomposition via MPI for distributing computational load across cluster nodes, with performance optimized through hybrid strategies combining MPI for inter-node communication and OpenMP for intra-node shared-memory parallelism \citep{alliataParallelizingDrugDiscovery2025,acunScalableMolecularDynamics2018,plimptonFastParallelAlgorithms1995}. Recent developments also include machine-learning-derived potentials through tools such as FitSNAP, achieving first-principles precision at reduced computational cost \citep{rohskopfFitSNAPAtomisticMachine2023}.

However, as the number of parallel processes increases, performance becomes progressively dependent on load balancing, communication latency, and hardware topology \citep{acunScalableMolecularDynamics2018,plimptonFastParallelAlgorithms1995}. The theoretical basis for this behavior was established by Amdahl, whose law predicts that the fraction of inherently serial or communication-bound work sets an absolute ceiling on parallel speedup regardless of the number of processors employed \citep{amdahlValiditySingleProcessor1967}; the present experiments provide an empirical quantification of that ceiling for coarse-grained LAMMPS workloads. In distributed architectures, the gains from parallelization compete directly with inter-process communication, synchronization, and data movement costs --- factors that can limit scalability at large scale. The operational efficiency of an HPC cluster depends on the harmony of the entire ecosystem, including memory bandwidth, parallel file systems \citep{kovatchOptimizingHighPerformanceComputing2020}, and the optimization of job scheduling and resource allocation \citep{kovatchOptimizingHighPerformanceComputing2020,ghediraDesignImplementationScalable2024}. The comparative evaluation of parallelization models is therefore essential for understanding the limits of performance, efficiency, and scalability in high-performance biomolecular simulations \citep{liHighperformanceComputingHealthcare2024,alliataParallelizingDrugDiscovery2025,thompsonLAMMPSFlexibleSimulation2022}.

\section{Methodology}\label{sec:methods}

The methodological framework of this study was structured to investigate the scalability behavior of the LAMMPS simulator in coarse-grained biomolecular applications executed in HPC environments. The motivation is grounded in the hypothesis that, in small- and medium-sized biomolecular systems, an excessive increase in the number of MPI ranks may lead to performance degradation due to the growth of inter-process communication costs, synchronization, and spatial fragmentation of the computational domain. Hybrid MPI+OpenMP approaches are therefore evaluated as a potential alternative to reduce inter-node communication overhead and improve resource usage on modern multicore and NUMA architectures.

\subsection{Computational environment and software stack}\label{sec:environment}

The experiments were carried out on the Lovelace cluster of the Centro Nacional de Processamento de Alto Desempenho em S\~ao Paulo (CENAPAD-SP), based on AMD EPYC 7662 nodes with 128 physical cores and 512 GB of RAM per node, interconnected by InfiniBand, with OpenMPI configured to use the UCX (Unified Communication X) communication layer. Each node provides a theoretical peak memory bandwidth of approximately 205 GB/s (eight-channel DDR4-3200), with an L3 cache totalling 256 MB distributed across eight Core Complex Dies (CCDs) at 32 MB per CCD. The simulator was LAMMPS version 30Mar2026, locally compiled with the system GCC 8.5.0 (Red Hat 8.5.0-18.0.2) and CMake 3.31.8; the runtime libraries from the \texttt{gcc/9.4.0} and \texttt{openmpi/4.1.1-gcc-9.4.0} modules were loaded by the PBS job scripts. The build enabled the CG-SPICA, KSPACE, MOLECULE, RIGID, and OPENMP packages required for coarse-grained simulations and hybrid parallelism.

Two parallelization models were used: pure MPI, in which each core executes one MPI rank, and hybrid MPI+OpenMP, in which multiple OpenMP threads are associated with each MPI rank to exploit shared-memory parallelism within each node. Process placement was controlled via OpenMPI directives (\texttt{--bind-to core} and \texttt{--map-by ppr:N:node:PE=T}), while the OpenMP environment used \texttt{OMP\_NUM\_THREADS}, \texttt{OMP\_PLACES=cores}, and \texttt{OMP\_PROC\_BIND=close} to keep threads close to their respective MPI rank, reducing remote memory access costs and synchronization overhead across the AMD EPYC CCDs (Core Complex Dies).

Executions were managed by the PBS scheduler. The par128 queue was used for single-node experiments and the paralela queue for multi-node experiments, ensuring all runs occurred on the same hardware family. To preserve reproducibility, all experiments used the same compiled binary, the same parallel libraries, and explicit initialization of the UCX environment in multi-node runs, avoiding inconsistencies in the MPI communication layer.

\subsection{Simulated biomolecular system}\label{sec:system}

The workload was built from the antimicrobial peptide Tritrpticin (PDB ID: 1D6X), a cationic peptide of 13 amino acid residues widely employed in peptide--membrane studies. The system was modeled in coarse-grained representation using SPICA force field version 2. Initial preparation was performed with the CHARMM-GUI SPICA CG Builder module, producing a DOPC lipid bilayer with explicit coarse-grained solvent, ions, and peptides distributed in the aqueous medium adjacent to the membrane. The atomistic-to-coarse-grained mapping was performed using the \texttt{map2cg} utility from the spica-tools package \citep{yusukemiyazakiSPICAgroupSpicatoolsV1002024}, and the peptide's structural restraints were defined through the Elastic Network Model (ENM) with secondary structure derived from DSSP.

The equilibrated simulation cell measured approximately $67.86 \times 67.86 \times 88.22$~\AA$^3$ at the start of the production NPT runs (as read from \texttt{eq.rest}) and contained 4354 coarse-grained beads. The anisotropic NPT barostat let the cell relax to approximately $66.47 \times 66.47 \times 79.87$~\AA$^3$ over the 5000 production steps. This is a small- to medium-sized workload, deliberately chosen to expose the breaking point of the PPPM solver under heavy communication pressure: at the largest core count evaluated (1024 ranks on 8 nodes), the workload distributes to roughly four particles per rank, an extreme boundary case intended to characterize the scalability ceiling for compact biomolecular systems rather than to demonstrate efficient strong-scaling on a workload sized to the hardware. Non-bonded interactions were modeled using the \texttt{lj/spica/coul/long} potential, combining Lennard-Jones forces with long-range electrostatic interactions treated through the \texttt{pppm/cg} method for charged coarse-grained systems.

\begin{figure}[pos=H]
  \centering
  \includegraphics[width=0.7\linewidth]{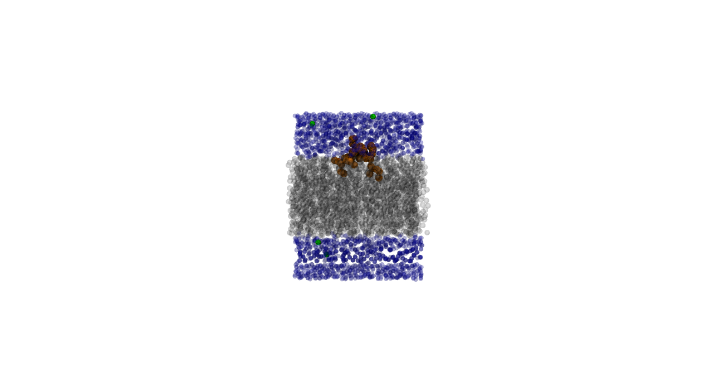}
  \caption{Coarse-grained biomolecular system based on the Tritrpticin peptide (1D6X) inserted into a hydrated DOPC lipid bilayer using the SPICA force field.}\label{fig:system}
\end{figure}

\subsection{Simulation parameters}\label{sec:params}

Simulations were conducted in the isothermal--isobaric (NPT) ensemble, with a Nose--Hoover thermostat and an anisotropic barostat compatible with coarse-grained lipid membrane systems. The \texttt{lj/spica/coul/long} potential used a 15~\AA{} cutoff for short-range interactions, while long-range electrostatics were handled by \texttt{pppm/cg} with a $10^{-5}$ error tolerance.

Neighbor lists were dynamically updated with a 2~\AA{} skin distance. The time integration step was 10 fs, and all executions ran 5000 production steps. Common restart files ensured that every configuration started from exactly the same thermodynamic state, so any observed performance differences could be attributed to the parallelization strategy and the HPC infrastructure rather than to variations in the molecular system.

\subsection{Experimental design}\label{sec:design}

The experiments were organized into two main groups: single-node and multi-node executions. The single-node group used one compute node with 128 physical cores to characterize intra-node scalability across different MPI$\times$OpenMP granularities while keeping the total core occupation constant. The multi-node group used 2, 4, and 8 nodes, totaling up to 1024 simultaneous cores, to investigate the effects of inter-node communication, NUMA topology, and distributed scalability. Each configuration was executed in triplicate using independent seeds (111, 222, 333) for the initialization of molecular velocities, yielding 54 main executions in total. Table~\ref{tbl:design} summarizes all configurations evaluated.

\begin{table}[pos=H]
\caption{Experimental configurations used in the LAMMPS scalability tests.}\label{tbl:design}
\begin{tabular*}{\tblwidth}{@{} LLLLL @{}}
\toprule
Nodes & MPI Ranks & OpenMP Threads & Total cores & Strategy \\
\midrule
1 & 128  & 1   & 128  & Pure MPI    \\
1 & 64   & 2   & 128  & Hybrid      \\
1 & 32   & 4   & 128  & Hybrid      \\
1 & 16   & 8   & 128  & Hybrid      \\
1 & 8    & 16  & 128  & Hybrid      \\
1 & 1    & 128 & 128  & Pure OpenMP \\
2 & 256  & 1   & 256  & Pure MPI    \\
2 & 128  & 2   & 256  & Hybrid      \\
2 & 64   & 4   & 256  & Hybrid      \\
2 & 32   & 8   & 256  & Hybrid      \\
4 & 512  & 1   & 512  & Pure MPI    \\
4 & 256  & 2   & 512  & Hybrid      \\
4 & 128  & 4   & 512  & Hybrid      \\
4 & 64   & 8   & 512  & Hybrid      \\
8 & 1024 & 1   & 1024 & Pure MPI    \\
8 & 512  & 2   & 1024 & Hybrid      \\
8 & 256  & 4   & 1024 & Hybrid      \\
8 & 128  & 8   & 1024 & Hybrid      \\
\bottomrule
\end{tabular*}
\end{table}

In all hybrid configurations, the total number of compute cores was kept constant within each multi-node scale, varying only the granularity between MPI processes and OpenMP threads. This approach made it possible to directly evaluate the impacts of hybrid decomposition on scalability behavior, synchronization, and distributed communication costs, ensuring that the differences observed between experiments were attributed exclusively to the parallelization strategies adopted.

\subsection{Performance metrics}\label{sec:metrics}

The primary metric was the total execution time of the production phase (Loop time) reported by LAMMPS, from which speedup and parallel efficiency were derived. Speedup was computed as the ratio between the execution time of the reference configuration and that of each evaluated configuration; parallel efficiency was obtained as the ratio between speedup and the relative number of compute resources employed. To attribute scalability behavior to specific computational components, the internal timing fractions reported by LAMMPS were analyzed for the Pair, Kspace, Comm, Neigh, Modify, and Output routines.

Hybrid executions were validated in two complementary ways. First, effective CPU utilization was measured via the time tool to confirm that OpenMP threads were actively running on the physical cores, ruling out silent fallback to serial execution. Second, numerical consistency between pure MPI and hybrid configurations was verified by comparing thermodynamic properties (temperature, pressure, box dimensions, and potential energy). Small numerical differences between configurations with distinct numbers of MPI ranks were interpreted as natural consequences of the non-commutativity of distributed floating-point sum operations, a phenomenon widely documented in parallel molecular dynamics \citep{plimptonFastParallelAlgorithms1995}. To support reproducibility and reduce the risk of transcription errors between the raw logs and the reported text, the supplementary repository accompanying this work includes a set of automated verification scripts that scrape the relevant infrastructure pages, parse the LAMMPS log files, regenerate the figures from the canonical CSV results, and cross-check every reference DOI against Crossref/DataCite; the reports produced by these scripts were used to confirm the numeric and bibliographic claims presented here.

\section{Results and discussion}\label{sec:results}

\subsection{Validation of hybrid parallelism}\label{sec:validation}

Prior to the scalability analysis, an experimental validation stage was carried out with the objective of verifying the correct operation of the hybrid MPI+OpenMP model used in the biomolecular experiments. This validation focused on two main aspects: (i) confirmation of the effective use of multiple OpenMP threads during the execution of the simulations, and (ii) verification of the numerical consistency of the trajectories produced by the hybrid executions compared to the pure MPI model.

The first stage measured effective CPU utilization. In every hybrid configuration analysed, utilization exceeded 100\%, confirming that multiple OpenMP threads were active in parallel rather than serialised. Single-node executions and the 8-node tests showed utilization above 96\%, close to the theoretical limit for the available cores.

Effective utilization peaked at approximately 99.3\% for the $128\times8$ configuration on 8 nodes, at the upper end of the range observed across all configurations.

Intermediate configurations distributed across 2 and 4 nodes showed reduced effective utilization, between approximately 77\% and 80\%. The section breakdown (Fig.~\ref{fig:breakdown}) shows that the Kspace+Comm fraction for these intermediate-node hybrid runs ranges from 71\% to 87\% of loop time, compared with 70\% for the $128\times8$ configuration on 8 nodes --- consistent with the observed efficiency gap.

The second stage of the validation focused on the numerical consistency of the hybrid executions. To this end, pure MPI ($128\times1$) and hybrid MPI+OpenMP ($128\times8$) executions were compared, keeping constant the global topology of MPI ranks and varying only the number of OpenMP threads associated with each process.

The comparison of thermodynamic properties demonstrated numerical equivalence between the two execution models. Temperature, pressure, simulation box dimensions, and potential energy presented bit-identical values across the three independent seeds evaluated after 5000 time integration steps under the NPT ensemble.

\begin{table}[pos=H]
\caption{Thermodynamic comparison between pure MPI and hybrid executions.}\label{tbl:thermo}
\begin{tabular*}{\tblwidth}{@{} LRRRRR @{}}
\toprule
Configuration & Temp (K) & Pxx (atm) & $L_x=L_y$ (\AA) & $L_z$ (\AA) & PotEng (kcal/mol) \\
\midrule
Pure MPI ($128\times1$)   & 298.37559 & $-7.5392328$ & 66.471948 & 79.868656 & $-26583.244$ \\
MPI+OpenMP ($128\times8$) & 298.37559 & $-7.5392328$ & 66.471948 & 79.868656 & $-26583.244$ \\
\bottomrule
\end{tabular*}
\end{table}

This bit-identical equivalence shows that the OpenMP execution path of LAMMPS preserves numerical results when the global spatial decomposition is held constant: only the per-rank thread count changes between the two runs.

Overall, the results of this stage validate both the numerical correctness and the computational effectiveness of the hybrid MPI+OpenMP model adopted in the subsequent experiments. In addition, the data obtained suggest that the scalability behavior observed in the coarse-grained biomolecular executions is strongly associated with the topological organization of the HPC hardware used, particularly with the effects of inter-node communication and NUMA affinity present in the AMD EPYC architecture.

\subsection{Scalability and execution time analysis}\label{sec:scalability}

\begin{figure}[pos=H]
  \centering
  \includegraphics[width=0.85\linewidth]{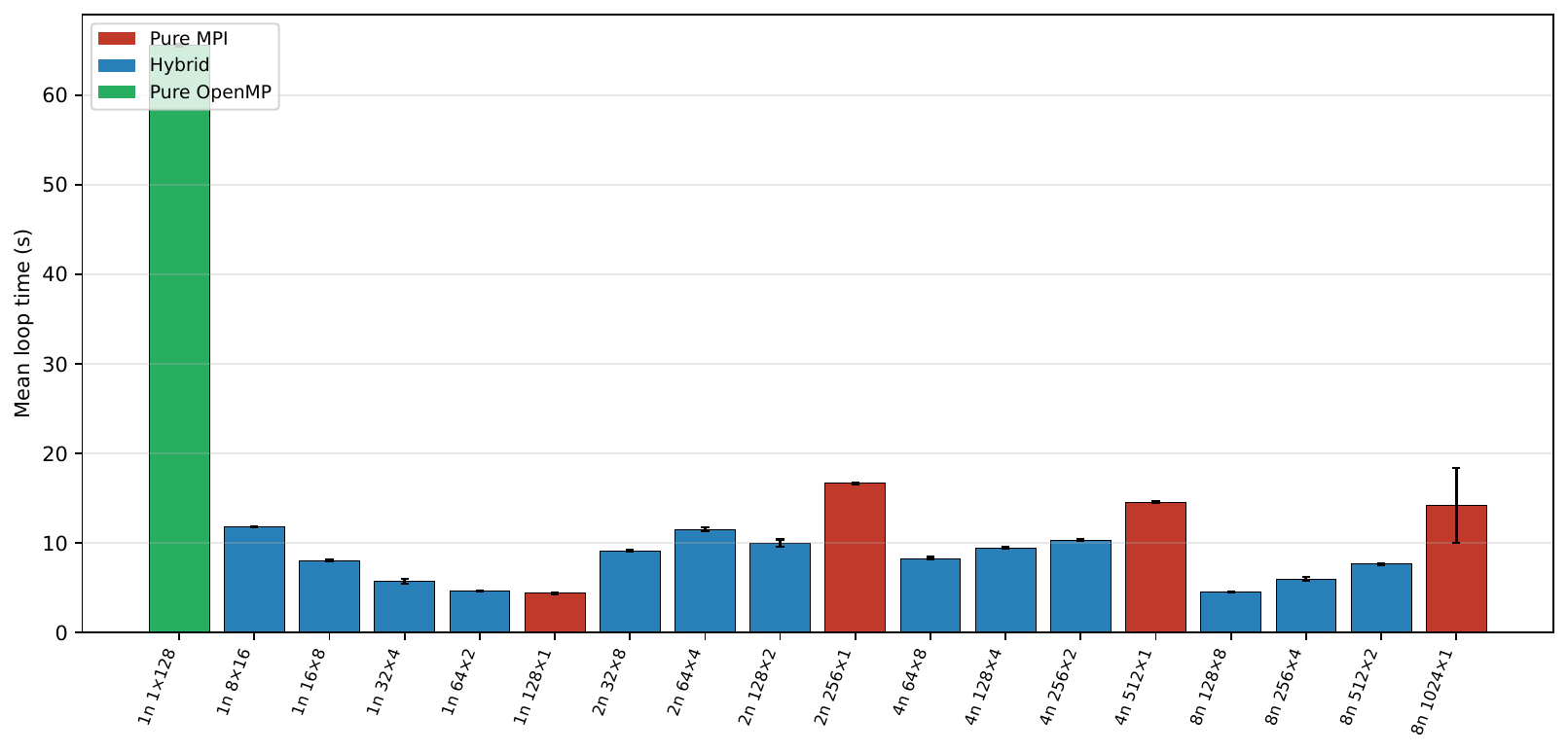}
  \caption{Comparison of mean execution time (loop time) between pure MPI, hybrid MPI+OpenMP, and pure OpenMP configurations across different multi-node scales.}\label{fig:loop}
\end{figure}

Figure~\ref{fig:loop} presents the behavior of the mean execution time (\textit{loop time}) for all experimental configurations evaluated, considering pure MPI, hybrid MPI+OpenMP, and pure OpenMP executions over different numbers of compute nodes.

\begin{figure}[pos=H]
  \centering
  \includegraphics[width=0.85\linewidth]{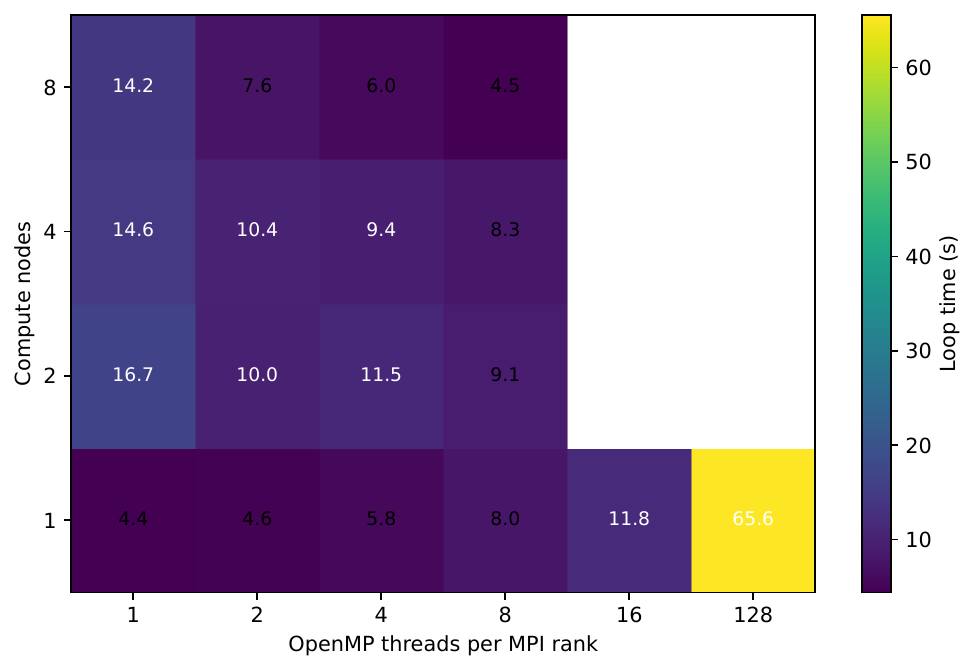}
  \caption{Heatmap of the mean execution time as a function of the number of compute nodes and the number of OpenMP threads per MPI rank.}\label{fig:heatmap}
\end{figure}

Figure~\ref{fig:heatmap} complements this analysis by presenting a heatmap of the execution time as a function of the number of nodes and the number of OpenMP threads per MPI rank. The existence of optimal regions of hybrid parallelization is clearly observed, particularly in the multi-node configurations with 4 and 8 OpenMP threads per MPI process.

The results show that the application's performance presented a strong dependence on the parallelization granularity adopted. In a single-node environment, the pure MPI configuration ($128\times1$) presented the lowest execution time observed across the entire experimental set, reaching approximately 4.40 s of loop time. Hybrid configurations with fewer MPI ranks and a greater number of OpenMP threads presented progressive performance degradation, culminating in the pure OpenMP configuration ($1\times128$), whose execution time exceeded 65 s.

On a single node, reducing MPI ranks in favour of OpenMP threads therefore degraded performance, from 4.40 s ($128\times1$) to over 65 s ($1\times128$). The section breakdown for the $1\times128$ case (Section~\ref{sec:breakdown}) shows the Pair routine consuming approximately 65\% of loop time, confirming that without MPI spatial decomposition the force-calculation phase does not parallelise efficiently across 128 threads on this workload.

However, the observed behavior changes substantially as the number of nodes increases. In the executions distributed across 2, 4, and 8 nodes, the hybrid configurations began to show better performance than the equivalent pure MPI executions. In particular, the hybrid $128\times8$ configuration executed on 8 nodes reached approximately 4.51 s, a value extremely close to the best single-node result ($128\times1$), even when using 1024 distributed cores.

On the other hand, the pure MPI executions showed progressive performance degradation with the increase in the number of nodes, reaching approximately 14.23 s in the $1024\times1$ configuration. This behavior evidences the growing impact of inter-process communication and global synchronization costs on multi-node architectures, particularly in coarse-grained biomolecular workloads with high frequency of information exchange between spatial domains.

Additionally, it is observed that certain intermediate hybrid configurations, such as $32\times8$ on 2 nodes and $64\times8$ on 4 nodes, presented relatively stable behavior, indicating that the reduction in the number of MPI ranks per node contributed to mitigating part of the communication overhead associated with distributed executions.

Overall, the results demonstrate that the efficiency of parallelization in biomolecular applications does not depend exclusively on the increase in the total number of compute cores, but mainly on the balance between spatial decomposition, MPI communication, and local threads parallelism.

\subsection{Speedup and parallel efficiency}\label{sec:speedup}

The analysis of speedup and parallel efficiency was conducted with the objective of investigating the scalability behavior of pure MPI and hybrid MPI+OpenMP executions as the total number of compute cores increased. For this evaluation, the best hybrid configurations observed at each number of nodes were compared with their respective equivalent pure MPI executions.

\begin{figure}[pos=H]
  \centering
  \includegraphics[width=0.85\linewidth]{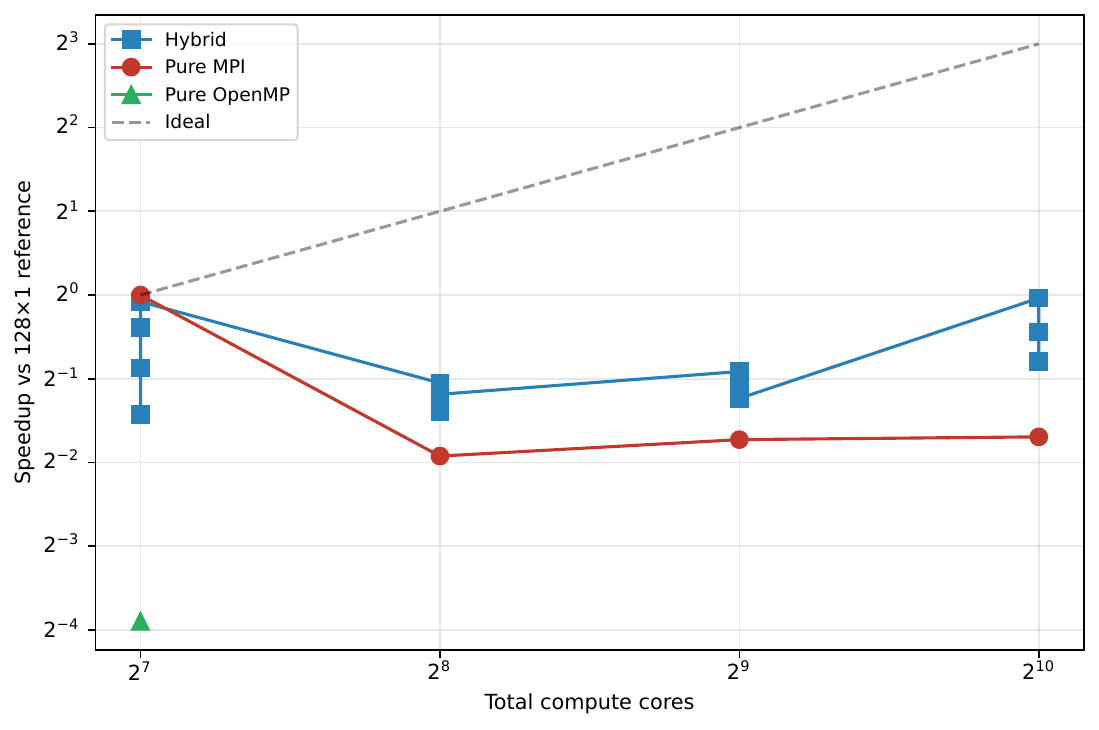}
  \caption{Comparison of speedup between pure MPI and hybrid MPI+OpenMP executions at different scales of distributed parallelization.}\label{fig:speedup}
\end{figure}

Figure~\ref{fig:speedup} presents the behavior of the \textit{speedup} obtained as a function of the total number of compute cores used in the experiments.

The results show that the hybrid MPI+OpenMP model presented consistently superior performance compared to pure MPI in multi-node environments. While the pure MPI executions suffered progressive degradation of speedup with the increase in the number of nodes, the hybrid configurations maintained relatively stable behavior, achieving the best results in the $64\times8$ and $128\times8$ configurations, executed respectively on 4 and 8 compute nodes.

It is observed that pure MPI showed a sharp performance drop after the transition from the single-node environment to distributed executions. The $256\times1$ configuration, executed on 2 nodes, presented speedup significantly lower than the $128\times1$ scenario, indicating that inter-process communication costs came to dominate a substantial part of the application's execution time.

This becomes most evident in the $1024\times1$ configuration, whose performance gain over smaller multi-node pure-MPI runs is marginal despite the increase in total cores. The cause is quantified in Section~\ref{sec:breakdown}: at 1024 ranks, Kspace+Comm together consume approximately 89\% of loop time, leaving little room for additional ranks to translate into speedup.

On the other hand, the hybrid configurations demonstrated greater robustness in the face of the increase in the number of nodes. The reduction in the number of MPI ranks per node made it possible to decrease the volume of distributed communication, preserving part of the computational locality through intra-node OpenMP parallelism. Consequently, the hybrid executions presented scalability behavior superior to pure MPI in practically all multi-node scenarios evaluated.

\begin{figure}[pos=H]
  \centering
  \includegraphics[width=0.85\linewidth]{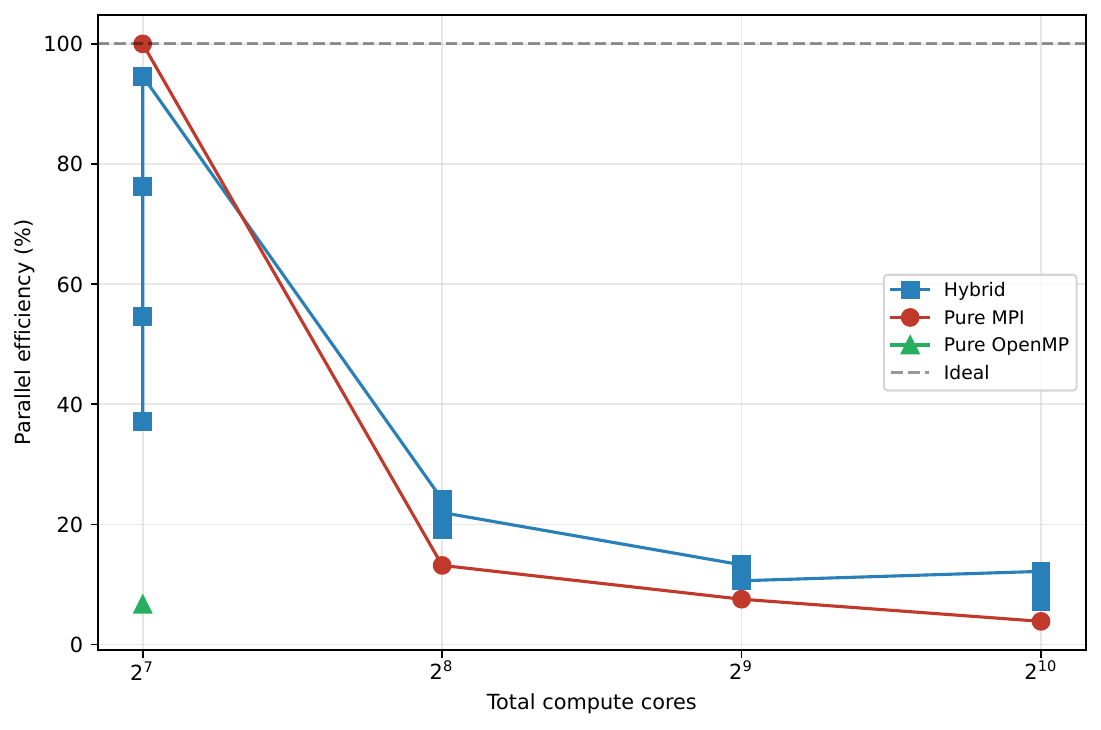}
  \caption{Comparison of parallel efficiency between pure MPI and hybrid MPI+OpenMP strategies on multi-node architectures.}\label{fig:efficiency}
\end{figure}

Figure~\ref{fig:efficiency} presents the parallel efficiency corresponding to the pure MPI and hybrid executions.

The efficiency results evidence a strong departure from ideal linear scaling in both parallelization models, an expected behavior for small- and medium-sized coarse-grained biomolecular workloads executed across a large number of distributed cores. Even so, the hybrid configurations maintained efficiency systematically superior to pure MPI, particularly in the executions involving 256, 512, and 1024 cores.

The low efficiency observed at the larger scales indicates that the biomolecular workload used has insufficient computational granularity to efficiently exploit thousands of MPI processes simultaneously. In this context, increasing parallelism comes to produce reduced marginal gains, while the costs associated with distributed communication become progressively dominant over the effective calculation of molecular forces.

Overall, the results demonstrate that hybrid MPI+OpenMP strategies constitute a more efficient alternative for coarse-grained biomolecular simulations in multi-node environments, particularly on modern many-core architectures, in which communication and synchronization costs represent a critical factor for the sustainability of parallel scalability.

\subsection{Analysis of the computational breakdown and communication costs}\label{sec:breakdown}

With the objective of investigating the factors responsible for the scalability behavior observed in the pure MPI and hybrid MPI+OpenMP executions, a detailed analysis of the internal temporal decomposition reported by LAMMPS was carried out. This evaluation made it possible to identify which computational components came to dominate the total execution time as the number of nodes and parallel processes increased.

\begin{figure}[pos=H]
  \centering
  \includegraphics[width=0.85\linewidth]{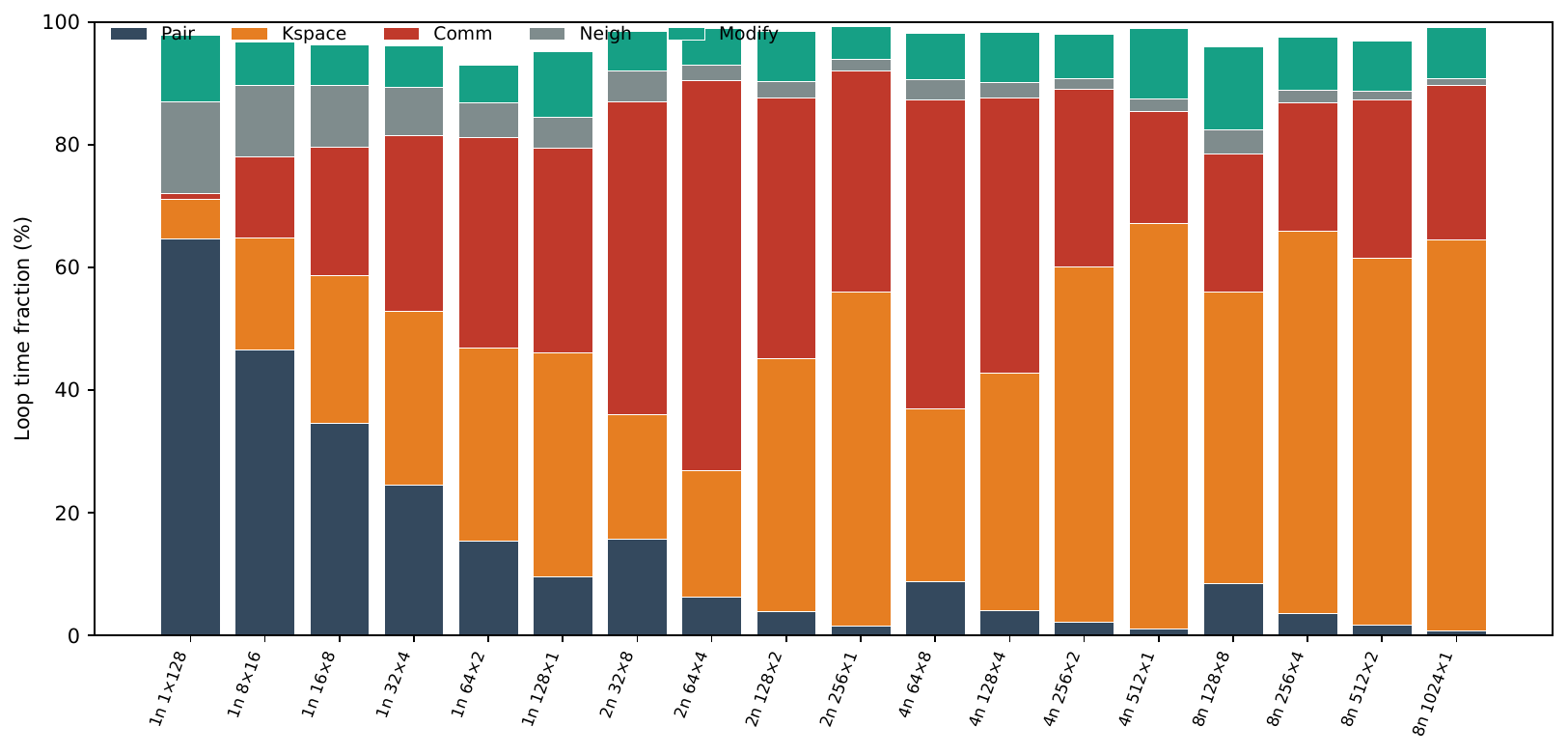}
  \caption{Percentage distribution of execution time among the main internal routines of LAMMPS for different parallelization strategies.}\label{fig:breakdown}
\end{figure}

Figure~\ref{fig:breakdown} presents the percentage breakdown of the main internal sections of LAMMPS for the best and worst configurations observed at each number of compute nodes.

The results show substantial differences between the hybrid and pure MPI executions. In the hybrid configurations considered most efficient ($32\times8$, $64\times8$, and $128\times8$), a relatively balanced distribution is observed between force calculation (Pair and Kspace) and communication (Comm). In contrast, the pure MPI executions presented strong predominance of communication and distributed electrostatic calculation routines, particularly in the $512\times1$ and $1024\times1$ configurations.

A memory-bound characterization further contextualizes these findings. The full simulation working set --- as reported by LAMMPS at approximately 203 MB per MPI rank in the single-rank reference run --- fits entirely within the 256 MB L3 cache of a single EPYC node. In distributed configurations with 8 MPI ranks per node, each rank's local subdomain occupies roughly 25 MB, remaining within the 32 MB L3 of a single CCD. Consequently, the pair force kernel operates predominantly from cache rather than from DRAM at both intra-CCD and intra-node granularities. The LAMMPS log for the single-rank reference run reports a per-rank memory allocation of 12.45 MB for the $128\times1$ case --- well below the 32 MB CCD-local L3 budget --- and 203 MB for the $1\times128$ case, which still fits within the 256 MB aggregate node L3. We do not claim a specific effective bandwidth figure here, since deriving one rigorously requires hardware-counter profiling (e.g.\ AMD uProf) that was not part of this study. The scalability ceiling observed in pure MPI executions therefore arises not from insufficient memory bandwidth per se, but from the communication and global synchronization overhead introduced by the PPPM electrostatic solver: the Kspace fraction grows from 36.5\% at $128\times1$ to 63.6\% at $1024\times1$, while the sum of Kspace and Comm reaches approximately 89\% of total loop time at the largest pure MPI scale, leaving less than 11\% of execution time for actual force computation. The best hybrid configuration at 8 nodes ($128\times8$) reduces this combined overhead to approximately 70\%, recovering a loop time of 4.51 s --- nearly matching the optimal single-node result of 4.40 s despite using 1024 distributed cores.

The pure OpenMP configuration ($1\times128$) presented behavior particularly distinct from the others, concentrating approximately 65\% of the total execution time in the Pair routine. This result evidences that the absence of MPI spatial decomposition severely compromises the efficient distribution of short-range force calculation on many-core architectures, producing intra-node computational saturation and significant performance degradation.

In the multi-node pure MPI executions, the growth of the time fraction associated with the Kspace and Comm routines became progressively dominant. This behavior indicates that the increase in the number of MPI \textit{ranks} substantially raised the cost of exchanging spatial halos, global synchronization, and distributed communication associated with the PPPM method employed to treat long-range electrostatic interactions.

\begin{figure}[pos=H]
  \centering
  \includegraphics[width=0.85\linewidth]{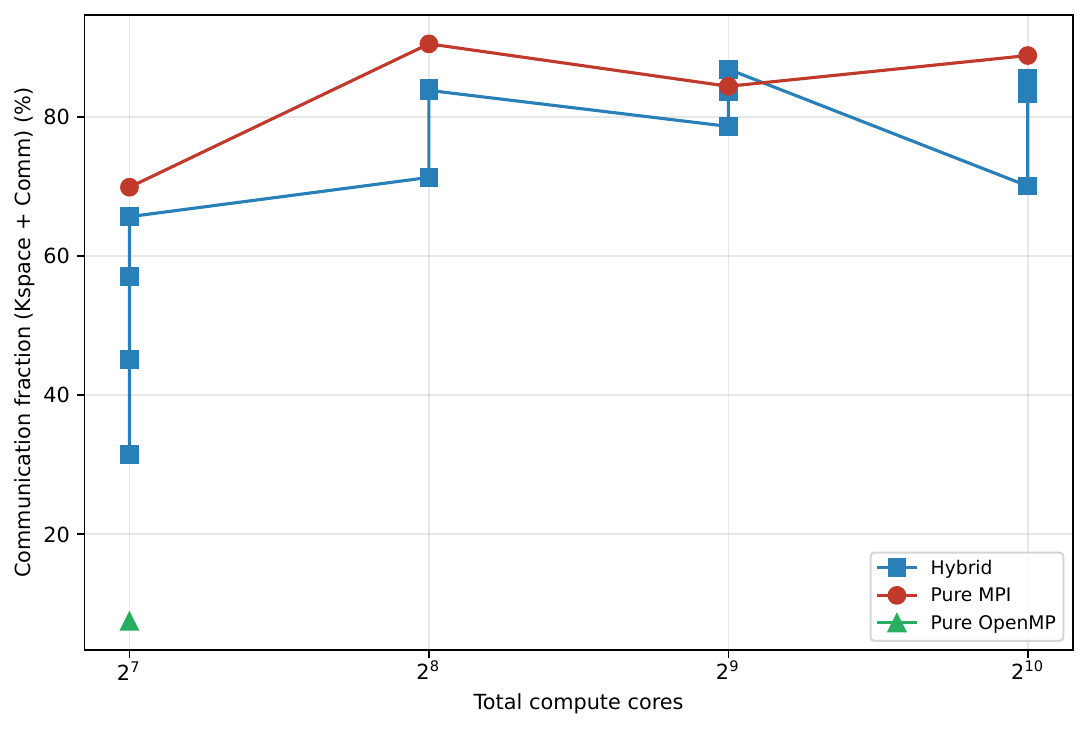}
  \caption{Growth of the relative cost of distributed communication (Kspace + Comm) with the increase in parallel scale.}\label{fig:kspacecomm}
\end{figure}

Figure~\ref{fig:kspacecomm} specifically presents the evolution of the routines related to communication (Kspace + Comm) as a function of the total number of compute cores used.

It is observed that the pure MPI executions reached approximately 90\% of the total time concentrated in communication-related routines at the largest scales analyzed. In contrast, the hybrid configurations maintained significantly lower values in several scenarios, particularly in the $32\times8$ and $128\times8$ executions, evidencing that the reduction in the number of MPI ranks per node directly contributed to mitigating the costs of distributed communication.

\begin{figure}[pos=H]
  \centering
  \includegraphics[width=\linewidth]{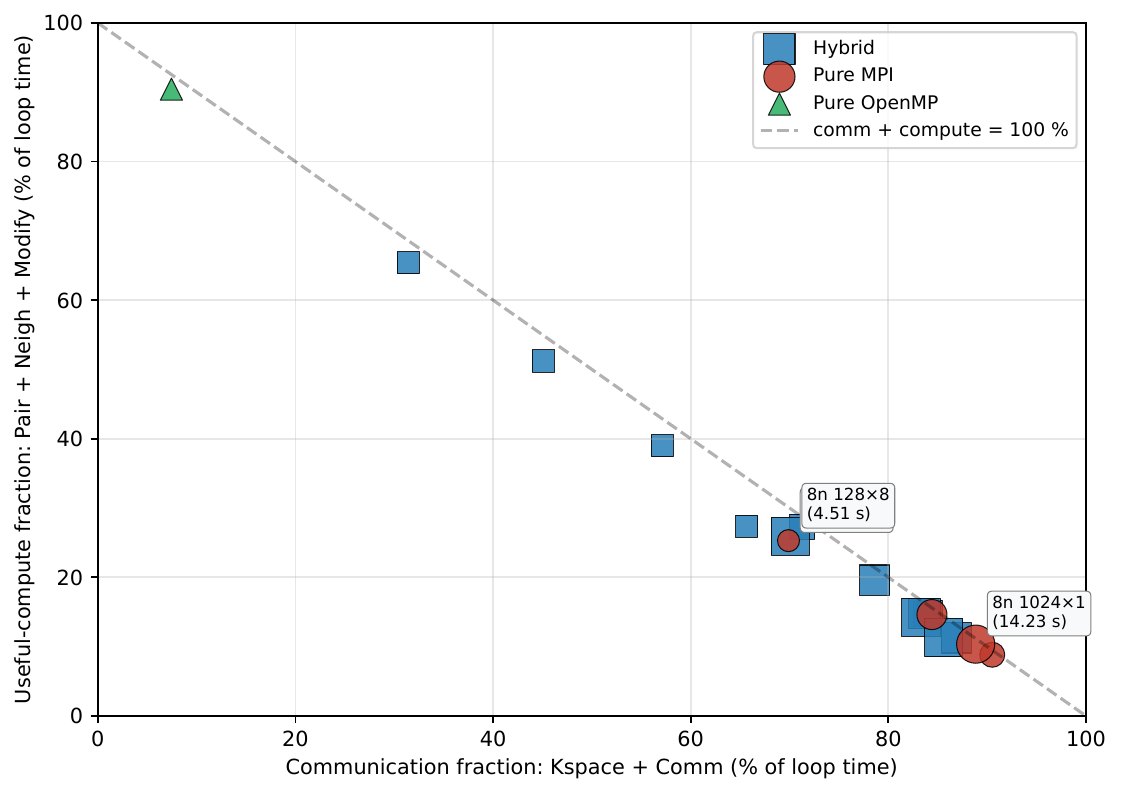}
  \caption{Trade-off between distributed communication overhead and time available for force computation. Each marker is one experimental configuration averaged across three seeds; marker size is proportional to the total number of compute cores. The dashed line indicates the geometric constraint that Kspace + Comm + useful-compute fractions cannot exceed 100\% of loop time.}\label{fig:roofline}
\end{figure}

Figure~\ref{fig:roofline} visualises this trade-off across all evaluated configurations. The x-axis is the communication fraction (Kspace + Comm) and the y-axis the useful-compute fraction (Pair + Neigh + Modify); marker size scales with the total number of cores. Configurations cluster along the diagonal because the remaining LAMMPS routines (Bond, Output, Other) contribute less than 2\% of loop time. Pure MPI executions migrate towards the bottom-right of the plot as scale increases, while the best hybrid configuration sits substantially higher on the compute axis even at 1024 cores.

This behavior reinforces the central hypothesis of this study: in small- and medium-sized coarse-grained biomolecular workloads, the indiscriminate growth in the number of MPI processes tends to produce excessive fragmentation of the spatial domain and a disproportionate increase in communication overhead. Consequently, a significant portion of the execution time ceases to be dedicated to the effective calculation of molecular interactions, coming to be consumed by synchronization and data exchange between distributed processes.

Overall, the analysis of the computational breakdown shows that the scalability limitation observed in the pure MPI executions does not arise exclusively from the processing capacity of the hardware, but mainly from the progressive growth of the cost of distributed communication in biomolecular applications executed on large-scale multi-node architectures.

\subsection{Integrated discussion of the results}\label{sec:discussion}

The experiments collectively show that the scalability behavior of coarse-grained biomolecular applications in HPC environments depends directly on the balance between spatial decomposition, parallelization granularity, and distributed communication cost. Increasing the total number of compute cores does not, by itself, translate into proportional performance gains for small- and medium-sized biomolecular workloads.

The three parallelization regimes evaluated behaved very differently. Pure MPI was most efficient on a single node but degraded sharply at multi-node scale, where Kspace and Comm routines came to dominate the loop time. Hybrid MPI+OpenMP was the most robust regime in distributed environments --- reducing the MPI rank count per node lowered the volume of inter-node communication and preserved CCD-local locality, exploiting the NUMA hierarchy of the AMD EPYC architecture. Pure OpenMP, by contrast, performed extremely poorly: without MPI spatial decomposition, the Pair routine concentrated the bulk of execution time on a single node, confirming that efficient parallelism for this workload requires both intra-node thread parallelism and an MPI-level decomposition of the simulation domain.

These observations also indicate that coarse-grained biomolecular workloads have computational characteristics distinct from HPC applications classically benchmarked for strong scaling. In relatively compact biomolecular systems, excessive fragmentation of the spatial domain produces subdomains too small to amortize the cost of synchronization and MPI halo exchange, leading to premature saturation of parallel scalability.

\begin{figure}[pos=H]
  \centering
  \includegraphics[width=0.85\linewidth]{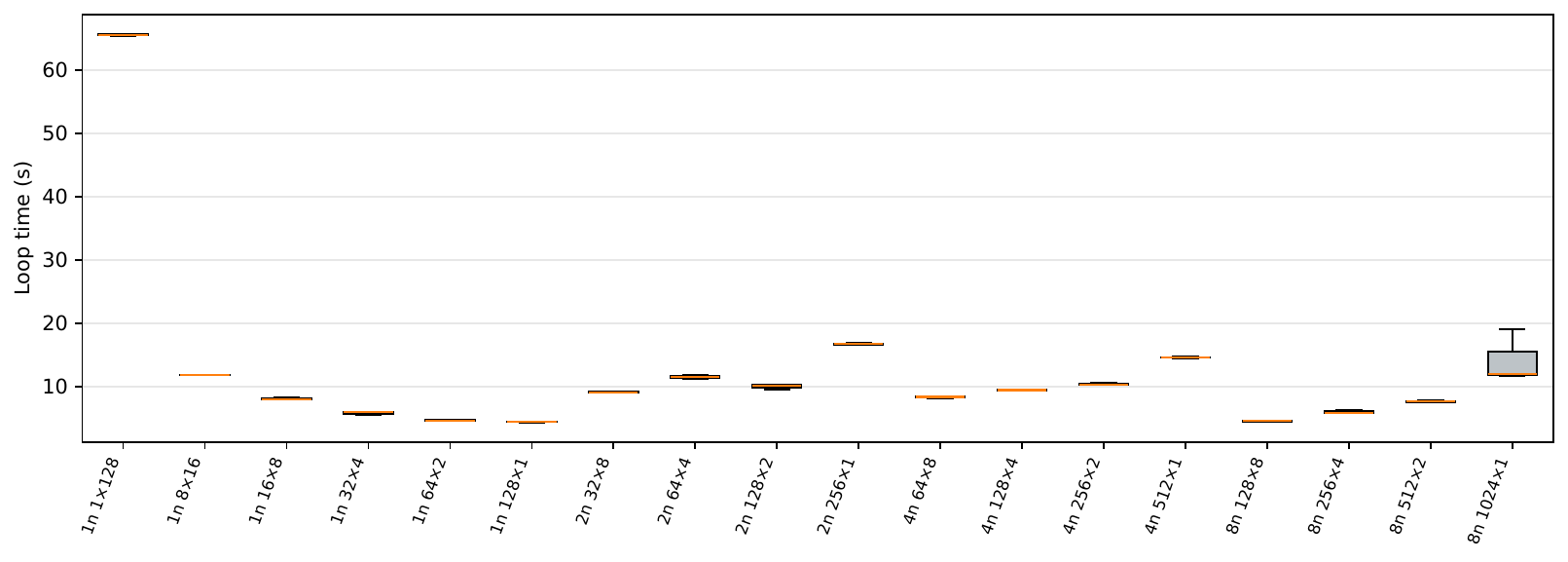}
  \caption{Percentage variability of the execution time across three independent seeds for each experimental configuration.}\label{fig:variance}
\end{figure}

Methodologically, the experiments demonstrated high statistical stability and numerical consistency between pure MPI and hybrid executions. Figure~\ref{fig:variance} presents the percentage dispersion of the execution time across the three independent seeds for each experimental configuration.

Most configurations presented variability below 10\% across seeds, reinforcing the stability of the measurements obtained and the thermodynamic equivalence between different parallelization strategies. The main exception was the $1024\times1$ configuration, whose variability exceeded 50\% --- consistent with the scalability saturation observed for pure MPI executions involving large numbers of distributed ranks, where small per-rank workloads amplify the relative impact of communication-time fluctuations.

Taken together, hybrid MPI+OpenMP strategies represent a more sustainable alternative for coarse-grained biomolecular simulations on modern multi-node architectures, particularly on many-core systems with complex NUMA organization. The performance of molecular dynamics applications should therefore be interpreted not from the total number of cores available but from the interplay between computational granularity, distributed communication, and hardware topology.

\section{Conclusion}\label{sec:conclusion}

This work investigated the scalability behavior of the LAMMPS simulator in coarse-grained biomolecular applications executed in High-Performance Computing environments, comparatively evaluating pure MPI and hybrid MPI+OpenMP parallelization strategies on multi-node architectures based on AMD EPYC processors. The experiments were conducted using the antimicrobial peptide Tritrpticin (PDB ID: 1D6X) as the biomolecular workload, allowing analysis of the impact of parallelization granularity on performance, parallel efficiency, and distributed communication cost metrics.

The results showed that pure MPI executions deliver excellent performance in single-node environments, particularly in the $128\times1$ configuration, which obtained the lowest execution time observed across the entire experimental set. However, as the number of compute nodes increased, progressive performance degradation and significant reduction of parallel efficiency were observed, evidencing a strong growth in the overhead associated with inter-process communication and distributed synchronization operations.

In contrast, the hybrid MPI+OpenMP configurations presented more robust behavior in multi-node environments, especially in the executions involving 4 and 8 compute nodes. The reduction in the number of MPI ranks per node made it possible to mitigate part of the distributed communication costs and to exploit the NUMA memory hierarchy of the architecture used more efficiently. As a consequence, hybrid configurations such as $64\times8$ and $128\times8$ presented performance significantly superior to the equivalent pure MPI executions at large scale.

The detailed analysis of the internal computational breakdown of LAMMPS revealed that the routines related to communication (Comm) and electrostatic treatment (Kspace) came to dominate the total execution time at the largest pure MPI scales, reaching approximately 90\% of the loop time in certain multi-node scenarios. This behavior evidences that, in small- and medium-sized coarse-grained biomolecular workloads, the indiscriminate increase in the number of MPI processes can produce premature saturation of parallel scalability due to excessive fragmentation of the spatial domain and the disproportionate growth of communication cost.

The experiments also demonstrated high statistical stability and numerical consistency among the executions performed, reinforcing the reliability of the experimental environment and the reproducibility of the measurements obtained. The low variability observed between independent seeds indicates that the performance differences identified arose effectively from the parallelization strategies evaluated, and not from random system fluctuations.

From the scientific and computational standpoint, the results reinforce that the efficiency of biomolecular applications in HPC environments does not depend exclusively on increasing the total number of cores available, but mainly on the balance between spatial decomposition, distributed communication, and intra-node parallelism. In this context, hybrid MPI+OpenMP strategies demonstrated greater computational sustainability for coarse-grained biomolecular simulations executed on modern many-core architectures.

Beyond the high-level scalability picture, the computational breakdown analysis identifies a concrete cause for the observed ceiling. As detailed in Section~\ref{sec:breakdown}, the PPPM electrostatic solver dominates loop time at large rank counts (Kspace + Comm approaching 89\% at 1024 ranks), and the simulation working set fits comfortably inside the CCD-local L3 cache of an EPYC node. Together these two observations indicate that the dominant bottleneck is the synchronization and global-reduction cost of the PPPM solver as the rank count grows, not raw memory bandwidth --- which is consistent with Amdahl's prediction for communication-bound work and is precisely what hybrid MPI+OpenMP configurations mitigate by keeping the per-node rank count low.

As future work, the extension of the analysis to larger-scale biomolecular systems is proposed, including complex lipid membranes and systems containing millions of coarse-grained particles, as well as the investigation of the impact of GPU accelerators and machine-learning-derived potentials on the scalability behavior of LAMMPS in heterogeneous HPC environments.

\section*{Acknowledgements}

This work was carried out within the research group led by Prof.\ F\'abio Andrijauskas at Universidade S\~ao Francisco (USF), which brings together undergraduate and master's researchers working on high-performance computing applied to biological data. The authors thank USF for the institutional support that made this collaboration possible. Computational resources were provided by the Centro Nacional de Processamento de Alto Desempenho em S\~ao Paulo (CENAPAD-SP, UNICAMP); the authors thank the CENAPAD-SP staff for assistance with the Lovelace cluster. This study was carried out under the research project \textit{T\'ecnicas de Computa\c{c}\~ao de Alto Desempenho para a Cria\c{c}\~ao de um Arcabou\c{c}o Voltado ao Processamento, Visualiza\c{c}\~ao e Compartilhamento de Dados Biol\'ogicos}, coordinated by Prof.\ Andrijauskas. The authors also acknowledge the LAMMPS and SPICA developer communities for maintaining the simulation engine and force field that made this work possible.

\section*{Declaration of competing interests}

The authors declare that they have no known competing financial interests or personal relationships that could have appeared to influence the work reported in this paper.

\section*{Data and software availability}

All LAMMPS input files, force field parameters, PBS job submission scripts, and analysis code used in this study are maintained in a Git repository (\url{https://github.com/DyeAllPies/LAMMPS_paper_1}), currently private and to be made publicly available upon publication; access can be provided to editors and reviewers on request. A citable archived snapshot will be deposited on Zenodo prior to final publication; the DOI will be updated here: [Zenodo DOI --- to be assigned]. The LAMMPS binary used was version 30Mar2026, compiled from the official source distribution (\url{https://www.lammps.org}) under the GNU GPL v2 license.

\section*{Declaration of generative AI and AI-assisted technologies in the writing process}

During the preparation of this manuscript, the authors used large language model tools --- including Claude (Anthropic, multiple versions including Opus and Sonnet families) and GPT-series models (OpenAI, versions 4.x and 5.x) --- to assist with language editing, text drafting and organization, literature search support, and generation of analysis and automation scripts. All AI-assisted content was reviewed, critically evaluated, and validated by the authors. The authors take full responsibility for the integrity, accuracy, and originality of all content in this publication.

\printcredits

\bibliographystyle{elsarticle-num}
\bibliography{references}

\begin{thebibliography}{10}
\expandafter\ifx\csname url\endcsname\relax
  \def\url#1{\texttt{#1}}\fi
\expandafter\ifx\csname urlprefix\endcsname\relax\def\urlprefix{URL }\fi
\expandafter\ifx\csname href\endcsname\relax
  \def\href#1#2{#2} \def\path#1{#1}\fi

\bibitem{liuApplyingHighperformanceComputing2016}
T.~Liu, D.~Lu, H.~Zhang, M.~Zheng, H.~Yang, Y.~Xu, C.~Luo, W.~Zhu, K.~Yu,
  H.~Jiang, Applying high-performance computing in drug discovery and molecular
  simulation, National Science Review 3~(1) (2016) 49--63.
\newblock \href {https://doi.org/10.1093/nsr/nww003}
  {\path{doi:10.1093/nsr/nww003}}.

\bibitem{hollingsworthMolecularDynamicsSimulation2018}
S.~A. Hollingsworth, R.~O. Dror, Molecular {{Dynamics Simulation}} for {{All}},
  Neuron 99~(6) (2018) 1129--1143.
\newblock \href {https://doi.org/10.1016/j.neuron.2018.08.011}
  {\path{doi:10.1016/j.neuron.2018.08.011}}.

\bibitem{acunScalableMolecularDynamics2018}
B.~Acun, D.~J. Hardy, L.~V. Kale, K.~Li, J.~C. Phillips, J.~E. Stone, Scalable
  molecular dynamics with {{NAMD}} on the {{Summit}} system, IBM Journal of
  Research and Development 62~(6) (2018) 4:1--4:9.
\newblock \href {https://doi.org/10.1147/jrd.2018.2888986}
  {\path{doi:10.1147/jrd.2018.2888986}}.

\bibitem{kutznerScalingGROMACSMolecular2025}
C.~Kutzner, V.~Mileti{\'c}, K.~Palacio~Rodr{\'i}guez, M.~Rampp, G.~Hummer,
  B.~L. {de Groot}, H.~Grubm{\"u}ller, Scaling of the {{GROMACS Molecular
  Dynamics Code}} to 65k {{CPU Cores}} on an {{HPC Cluster}}, Journal of
  Computational Chemistry 46~(5) (Feb. 2025).
\newblock \href {https://doi.org/10.1002/jcc.70059}
  {\path{doi:10.1002/jcc.70059}}.

\bibitem{kochHPCMedicalField2023}
M.~Koch, C.~Arlandini, G.~Antonopoulos, A.~Baretta, P.~Beaujean, G.~J. Bex,
  M.~E. Biancolini, S.~Celi, E.~Costa, L.~Drescher, V.~Eleftheriadis, N.~A.
  Fadel, A.~Fink, F.~Galbiati, I.~Hatzakis, G.~Hompis, N.~Lewandowski,
  A.~Memmolo, C.~Mensch, D.~Obrist, V.~Paneta, P.~Papadimitroulas,
  K.~Petropoulos, S.~Porziani, G.~Savvidis, K.~Sethia, P.~Strakos,
  P.~Svobodova, E.~Vignali, {{HPC}}+ in the medical field: {{Overview}} and
  current examples, Technology and Health Care 31~(4) (2023) 1509--1523.
\newblock \href {https://doi.org/10.3233/thc-229015}
  {\path{doi:10.3233/thc-229015}}.

\bibitem{stephensBigDataAstronomical2015}
Z.~D. Stephens, S.~Y. Lee, F.~Faghri, R.~H. Campbell, C.~Zhai, M.~J. Efron,
  R.~Iyer, M.~C. Schatz, S.~Sinha, G.~E. Robinson, Big {{Data}}:
  {{Astronomical}} or {{Genomical}}?, PLOS Biology 13~(7) (2015) e1002195.
\newblock \href {https://doi.org/10.1371/journal.pbio.1002195}
  {\path{doi:10.1371/journal.pbio.1002195}}.

\bibitem{lightbodyReviewApplicationsHighthroughput2019}
G.~Lightbody, V.~Haberland, F.~Browne, L.~Taggart, H.~Zheng, E.~Parkes, J.~K.
  Blayney, Review of applications of high-throughput sequencing in personalized
  medicine: Barriers and facilitators of future progress in research and
  clinical application, Briefings in Bioinformatics 20~(5) (2019) 1795--1811.
\newblock \href {https://doi.org/10.1093/bib/bby051}
  {\path{doi:10.1093/bib/bby051}}.

\bibitem{deanMapReduceSimplifiedData2008}
J.~Dean, S.~Ghemawat, {{MapReduce}}: Simplified data processing on large
  clusters, Communications of the ACM 51~(1) (2008) 107--113.
\newblock \href {https://doi.org/10.1145/1327452.1327492}
  {\path{doi:10.1145/1327452.1327492}}.

\bibitem{gothEricBrewerChange2010}
G.~Goth, Eric {{Brewer}}: Change agent, Communications of the ACM 53~(7) (2010)
  24--24.
\newblock \href {https://doi.org/10.1145/1785414.1785425}
  {\path{doi:10.1145/1785414.1785425}}.

\bibitem{kovatchOptimizingHighPerformanceComputing2020}
P.~Kovatch, L.~Gai, H.~M. Cho, E.~Fluder, D.~Jiang, Optimizing
  {{High-Performance Computing Systems}} for {{Biomedical Workloads}}, in: 2020
  {{IEEE International Parallel}} and {{Distributed Processing Symposium
  Workshops}} ({{IPDPSW}}), IEEE, 2020, pp. 183--192.
\newblock \href {https://doi.org/10.1109/ipdpsw50202.2020.00040}
  {\path{doi:10.1109/ipdpsw50202.2020.00040}}.

\bibitem{liHighperformanceComputingHealthcare2024}
J.~Li, S.~Wang, S.~Rudinac, A.~Osseyran, High-performance computing in
  healthcare: {{An}} automatic literature analysis perspective, Journal of Big
  Data 11~(1) (May 2024).
\newblock \href {https://doi.org/10.1186/s40537-024-00929-2}
  {\path{doi:10.1186/s40537-024-00929-2}}.

\bibitem{mashayakRelativeEntropyOptimizationDriven2015}
S.~Y. Mashayak, M.~N. Jochum, K.~Koschke, N.~R. Aluru, V.~R{\"u}hle,
  C.~Junghans, Relative {{Entropy}} and {{Optimization-Driven Coarse-Graining
  Methods}} in {{VOTCA}}, PLOS ONE 10~(7) (2015) e0131754.
\newblock \href {https://doi.org/10.1371/journal.pone.0131754}
  {\path{doi:10.1371/journal.pone.0131754}}.

\bibitem{alliataParallelizingDrugDiscovery2025}
P.~R. Alliata, D.~Rubaga, D.~Kumlin, A.~Puliga, Parallelizing {{Drug
  Discovery}}: {{HPC Pipelines}} for {{Alzheimer}}'s {{Molecular Docking}} and
  {{Simulation}} (2025).
\newblock \href {https://doi.org/10.48550/ARXIV.2509.00937}
  {\path{doi:10.48550/ARXIV.2509.00937}}.

\bibitem{marrinkPerspectiveMartiniModel2013}
S.~J. Marrink, D.~P. Tieleman, Perspective on the {{Martini}} model, Chemical
  Society Reviews 42~(16) (2013) 6801.
\newblock \href {https://doi.org/10.1039/c3cs60093a}
  {\path{doi:10.1039/c3cs60093a}}.

\bibitem{braunBestPracticesFoundations2018}
E.~Braun, J.~Gilmer, H.~B. Mayes, D.~L. Mobley, J.~I. Monroe, S.~Prasad, D.~M.
  Zuckerman, Best {{Practices}} for {{Foundations}} in {{Molecular
  Simulations}} [{{Article}} v1.0], Living Journal of Computational Molecular
  Science 1~(1) (2018) 5957.
\newblock \href {https://doi.org/10.33011/livecoms.1.1.5957}
  {\path{doi:10.33011/livecoms.1.1.5957}}.

\bibitem{georgievaProteinConformationalDynamics2020}
E.~R. Georgieva, Protein {{Conformational Dynamics}} upon {{Association}} with
  the {{Surfaces}} of {{Lipid Membranes}} and {{Engineered Nanoparticles}}:
  {{Insights}} from {{Electron Paramagnetic Resonance Spectroscopy}}, Molecules
  25~(22) (2020) 5393.
\newblock \href {https://doi.org/10.3390/molecules25225393}
  {\path{doi:10.3390/molecules25225393}}.

\bibitem{ramirezModelingContextSpecificEMT2020}
D.~Ramirez, V.~Kohar, M.~Lu, Toward {{Modeling Context-Specific EMT Regulatory
  Networks Using Temporal Single Cell RNA-Seq Data}}, Frontiers in Molecular
  Biosciences 7 (Apr. 2020).
\newblock \href {https://doi.org/10.3389/fmolb.2020.00054}
  {\path{doi:10.3389/fmolb.2020.00054}}.

\bibitem{herberClickChemistrymediatedBiotinylation2018}
J.~Herber, J.~Njavro, R.~Feederle, U.~Schepers, U.~C. M{\"u}ller, S.~Br{\"a}se,
  S.~A. M{\"u}ller, S.~F. Lichtenthaler, Click {{Chemistry-mediated
  Biotinylation Reveals}} a {{Function}} for the {{Protease BACE1}} in
  {{Modulating}} the {{Neuronal Surface Glycoproteome}}, Molecular \&amp;
  Cellular Proteomics 17~(8) (2018) 1487--1501.
\newblock \href {https://doi.org/10.1074/mcp.ra118.000608}
  {\path{doi:10.1074/mcp.ra118.000608}}.

\bibitem{adrianFLASHEffectDepends2019}
G.~Adrian, E.~Konradsson, M.~Lempart, S.~B{\"a}ck, C.~Ceberg, K.~Petersson, The
  {{FLASH}} effect depends on oxygen concentration, The British Journal of
  Radiology 93~(1106) (Dec. 2019).
\newblock \href {https://doi.org/10.1259/bjr.20190702}
  {\path{doi:10.1259/bjr.20190702}}.

\bibitem{seoSPICAForceField2018}
S.~Seo, W.~Shinoda, {{SPICA Force Field}} for {{Lipid Membranes}}: {{Domain
  Formation Induced}} by {{Cholesterol}}, Journal of Chemical Theory and
  Computation 15~(1) (2018) 762--774.
\newblock \href {https://doi.org/10.1021/acs.jctc.8b00987}
  {\path{doi:10.1021/acs.jctc.8b00987}}.

\bibitem{monticelliForceFieldsClassical2012}
L.~Monticelli, D.~P. Tieleman, Force {{Fields}} for {{Classical Molecular
  Dynamics}}, in: Biomolecular {{Simulations}}, Humana Press, 2012, pp.
  197--213.
\newblock \href {https://doi.org/10.1007/978-1-62703-017-5_8}
  {\path{doi:10.1007/978-1-62703-017-5_8}}.

\bibitem{drorBiomolecularSimulationComputational2012}
R.~O. Dror, R.~M. Dirks, J.~Grossman, H.~Xu, D.~E. Shaw, Biomolecular
  {{Simulation}}: {{A Computational Microscope}} for {{Molecular Biology}},
  Annual Review of Biophysics 41~(1) (2012) 429--452.
\newblock \href {https://doi.org/10.1146/annurev-biophys-042910-155245}
  {\path{doi:10.1146/annurev-biophys-042910-155245}}.

\bibitem{lindorff-larsenSystematicValidationProtein2012}
K.~{Lindorff-Larsen}, P.~Maragakis, S.~Piana, M.~P. Eastwood, R.~O. Dror, D.~E.
  Shaw, Systematic {{Validation}} of {{Protein Force Fields}} against
  {{Experimental Data}}, PLoS ONE 7~(2) (2012) e32131.
\newblock \href {https://doi.org/10.1371/journal.pone.0032131}
  {\path{doi:10.1371/journal.pone.0032131}}.

\bibitem{gelpiMolecularDynamicsSimulations2015}
J.~Gelpi, A.~Hospital, R.~Go{\~n}i, M.~Orozco, Molecular dynamics simulations:
  Advances and applications, Advances and Applications in Bioinformatics and
  Chemistry (2015) 37\href {https://doi.org/10.2147/aabc.s70333}
  {\path{doi:10.2147/aabc.s70333}}.

\bibitem{vanommeslaegheCHARMMGeneralForce2009}
K.~Vanommeslaeghe, E.~Hatcher, C.~Acharya, S.~Kundu, S.~Zhong, J.~Shim,
  E.~Darian, O.~Guvench, P.~Lopes, I.~Vorobyov, A.~D. Mackerell, {{CHARMM}}
  general force field: {{A}} force field for drug-like molecules compatible
  with the {{CHARMM}} all-atom additive biological force fields, Journal of
  Computational Chemistry 31~(4) (2009) 671--690.
\newblock \href {https://doi.org/10.1002/jcc.21367}
  {\path{doi:10.1002/jcc.21367}}.

\bibitem{nochezAcuteLateonsetOptic2009}
Y.~Nochez, S.~Arsene, N.~Gueguen, A.~Chevrollier, M.~Ferr{\'e}, V.~Guillet,
  et~al., Acute and late-onset optic atrophy due to a novel {{OPA1}} mutation
  leading to a mitochondrial coupling defect, Molecular Vision 15 (2009)
  598--608.

\bibitem{righettiConstrainedAmorphousInterphase2019}
M.~C. Righetti, L.~Aliotta, N.~Mallegni, M.~Gazzano, E.~Passaglia, P.~Cinelli,
  A.~Lazzeri, Constrained {{Amorphous Interphase}} and {{Mechanical
  Properties}} of {{Poly}}(3-{{Hydroxybutyrate-co-3-Hydroxyvalerate}}),
  Frontiers in Chemistry 7 (Nov. 2019).
\newblock \href {https://doi.org/10.3389/fchem.2019.00790}
  {\path{doi:10.3389/fchem.2019.00790}}.

\bibitem{plimptonFastParallelAlgorithms1995}
S.~Plimpton, Fast {{Parallel Algorithms}} for {{Short-Range Molecular
  Dynamics}}, Journal of Computational Physics 117~(1) (1995) 1--19.
\newblock \href {https://doi.org/10.1006/jcph.1995.1039}
  {\path{doi:10.1006/jcph.1995.1039}}.

\bibitem{essmannSmoothParticleMesh1995}
U.~Essmann, L.~Perera, M.~L. Berkowitz, T.~Darden, H.~Lee, L.~G. Pedersen, A
  smooth particle mesh {{Ewald}} method, The Journal of Chemical Physics
  103~(19) (1995) 8577--8593.
\newblock \href {https://doi.org/10.1063/1.470117}
  {\path{doi:10.1063/1.470117}}.

\bibitem{thompsonLAMMPSFlexibleSimulation2022}
A.~P. Thompson, H.~M. Aktulga, R.~Berger, D.~S. Bolintineanu, W.~M. Brown,
  P.~S. Crozier, P.~J. {in 't Veld}, A.~Kohlmeyer, S.~G. Moore, T.~D. Nguyen,
  R.~Shan, M.~J. Stevens, J.~Tranchida, C.~Trott, S.~J. Plimpton, {{LAMMPS}} -
  a flexible simulation tool for particle-based materials modeling at the
  atomic, meso, and continuum scales, Computer Physics Communications 271
  (2022) 108171.
\newblock \href {https://doi.org/10.1016/j.cpc.2021.108171}
  {\path{doi:10.1016/j.cpc.2021.108171}}.

\bibitem{rohskopfFitSNAPAtomisticMachine2023}
A.~Rohskopf, C.~Sievers, N.~Lubbers, M.~A. Cusentino, J.~Goff, J.~Janssen,
  M.~McCarthy, D.~M. {de Oca Zapiain}, S.~Nikolov, K.~Sargsyan, D.~Sema,
  E.~Sikorski, L.~Williams, A.~P. Thompson, M.~A. Wood, {{FitSNAP}}:
  {{Atomistic}} machine learning with {{LAMMPS}}, Journal of Open Source
  Software 8~(84) (2023) 5118.
\newblock \href {https://doi.org/10.21105/joss.05118}
  {\path{doi:10.21105/joss.05118}}.

\bibitem{amdahlValiditySingleProcessor1967}
G.~M. Amdahl, Validity of the single processor approach to achieving large
  scale computing capabilities, in: Proceedings of the {{April}} 18-20, 1967,
  Spring Joint Computer Conference on - {{AFIPS}} '67 ({{Spring}}), {{AFIPS}}
  '67 ({{Spring}}), ACM Press, 1967, p. 483.
\newblock \href {https://doi.org/10.1145/1465482.1465560}
  {\path{doi:10.1145/1465482.1465560}}.

\bibitem{ghediraDesignImplementationScalable2024}
K.~Ghedira, O.~Khamessi, C.~Hkimi, S.~Kamoun, N.~Dhamer, K.~Daassi,
  W.~Ben~Salah, H.~Othman, W.~Belhadj, Y.~Ghorbal, Design and implementation of
  a scalable high-performance computing ({{HPC}}) cluster for omics data
  analysis: Achievements, challenges and recommendations in {{LMICs}},
  GigaScience 13 (2024).
\newblock \href {https://doi.org/10.1093/gigascience/giae060}
  {\path{doi:10.1093/gigascience/giae060}}.

\bibitem{yusukemiyazakiSPICAgroupSpicatoolsV1002024}
Y.~Miyazaki, Y.~Teppei, W.~Shinoda, {onefive13}, {{SPICA-group}}/spica-tools:
  V1.0.0, Zenodo (Feb. 2024).
\newblock \href {https://doi.org/10.5281/ZENODO.10611578}
  {\path{doi:10.5281/ZENODO.10611578}}.

\end{thebibliography}

\end{document}